\documentclass[
reprint,
superscriptaddress,
amsmath,amssymb,
aps,
prb,
]{revtex4-2}

\usepackage{graphicx}
\usepackage{dcolumn}
\usepackage{xcolor}
\usepackage{bm}
\usepackage[upgreek]{mathastext}
\usepackage[section]{placeins}
\usepackage{hyperref}
\hypersetup{colorlinks=true, citecolor=blue, urlcolor=blue, linkcolor=blue}
\usepackage{titlesec}
\usepackage{needspace}

\begin{document}
\title{Exploring  Magnetism of Lead-free Halide Double Perovskites: A High-Throughput First-Principles Study}
\author{Utkarsh Singh}
\email{utkarsh.singh[at]liu.se}
\affiliation{%
Theoretical Physics Division, \\
Department of Physics, Chemistry and Biology (IFM),
Link\"{o}ping University, SE-581 83, Link\"{o}ping, Sweden
}%
\author{Johan Klarbring}
\affiliation{%
Theoretical Physics Division, \\
Department of Physics, Chemistry and Biology (IFM),
Link\"{o}ping University, SE-581 83, Link\"{o}ping, Sweden
}%
\author{Igor A. Abrikosov}
\affiliation{%
Theoretical Physics Division, \\
Department of Physics, Chemistry and Biology (IFM),
Link\"{o}ping University, SE-581 83, Link\"{o}ping, Sweden
}%
\author{Sergei I. Simak}
\affiliation{%
Theoretical Physics Division, \\
Department of Physics, Chemistry and Biology (IFM),
Link\"{o}ping University, SE-581 83, Link\"{o}ping, Sweden
}%
\affiliation{%
Department of Physics and Astronomy,
\\
Uppsala University, SE-75120 Uppsala, Sweden
}%

\date{\today}

\begin{abstract}
    We have performed a comprehensive, first-principles high-throughput study of the magnetic properties of halide double perovskites, $Cs_2BB^\prime Cl_6$, with magnetic ions occupying one or both B and B$^\prime$ sites. Our findings indicate a general tendency for these materials to exhibit antiferromagnetic ordering with low N\'eel temperatures. At the same time, we reveal a few potential candidates that predicted to be ferromagnetic with relatively high Curie temperatures. Achieving ferromagnetic coupling might be feasible via simultaneously alloying at B and B$^\prime$ sites with magnetic 3d and non-magnetic 5d ions. With this approach, we discover that $Cs_2HgCrCl_6$, $Cs_2AgNiCl_6$ and $Cs_2AuNiCl_6$ have high Curie temperatures relative to their peers, with the latter two exhibiting half metallic behaviour. Further, this study illuminates the underpinning mechanism of magnetic exchange interactions in halide double perovskites, enabling a deeper understanding of their magnetic behaviour. Our findings, especially the discovery of the compounds with robust half-metallic properties and high Curie temperatures holds promise for potential applications in the field of spintronics.
\end{abstract}


\maketitle

\section{\label{sec:intro}Introduction\protect}

Over the past decade, research on metal halide perovskites has experienced a consistent growth, establishing itself as one of the most dynamic fields within materials science \cite{Jena2019}. While the investigation was initially focused on materials containing lead, several lead-free variations have emerged as intriguing alternatives \cite{Ning2019}. These subsets primarily aim to overcome the limitations associated not only with the toxicity of lead but also with the poor thermodynamic stability. One particular group of materials that has surfaced as an alternative in recent years is lead-free halide double perovskites (HDPs), wherein the Pb$^{+2}$ ions are substituted with a combination of monovalent and trivalent cations \cite{Igbari2019}.

\medskip

Lead-free halide double perovskites (LFHDPs) $A_2BB^\prime Cl_6$ are composed of A $\in$ \{Cs, Rb\}, B $\in$ \{+1 elements\}, B$^\prime$ $\in$ \{+3 elements\}, and X $\in$ \{Cl, Br, I\}. They present non-toxic and stable alternatives to typical lead-based $ABX_3$ perovskites with A $\in$ \{MA$^+$, FA$^+$, Cs\}, B = 2+ toxic element Pb, and X $\in$ \{F, Cl, Br, I\}. Here MA stands for methylammonium and FA stands for formamidinium.

\medskip

The combination of (+1, +3) ions on (B, B$^\prime$) sublattices in LFHDPs provides a larger chemical space in comparison to typical APbX$_{3}$ perovskites, making halide double perovskites, potentially, a class of multifunctional materials \cite{Lei2021, Zhu2022} similar to Heusler compounds \cite{Graf2011} and their oxide counterparts \cite{Vasala2015}. One facet of this is applications based on their magnetic properties, which require substitution of either or both B and B$^\prime$ sites with elements having unpaired d - electrons.\par

Recently, several studies have shown that Fe$^{+3}$ can be incoorporated in the double perovskite structure in the B$^\prime$ sublattice when the B sublattice is populated with s-block element Na \cite{Zhang2023, Armer2023} and d-block element, Ag \cite{Yin2020}. However, unlike their oxide counterparts, for LFHDPs, this has mainly been explored in the context of improving the performance of optoelectronic devices \cite{Fuxiang2021, Han2021}, leaving out many possible combinations of elements which can populate the structure. We therefore find it important to examine, qualitatively and quantitatively, the nature of magnetic interactions in halide double perovskites.

\medskip

\begin{figure*}[!htb]
    \includegraphics[scale = 0.20]{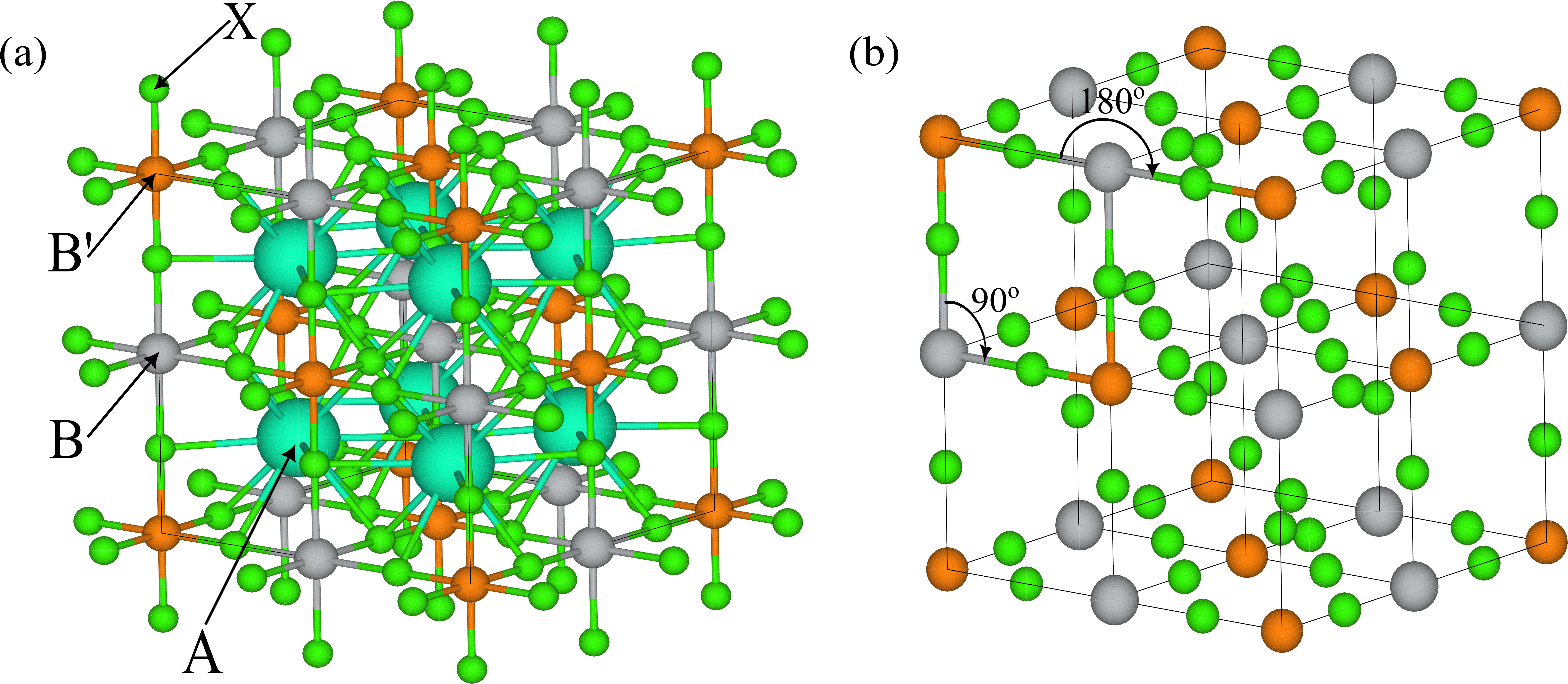}
    \caption{Illustrtation of (a) Conventional unit cell for halide double perovskite
structure, $A_2BB^\prime X_6$, A = Cs, Rb, B, B$^\prime$ $\in$ {+1, +3} or {+2, +2} elements, X = Cl, Br, I (b) Exchange interaction path, B$^\prime$ - X - B - X - B$^\prime$ where $\angle $ X - B - X = 90$^\circ$ or 180$^\circ$ for first and second coordination shells respectively.}
    \label{fig:structure}
\end{figure*}

The double perovskite structure, shown in Figure \ref{fig:structure} (a), can be described as two interpenetrating face-centered cubic (fcc) B and B$^\prime$ sublattices, with the species A occupying a simple cubic lattice. This structure prohibits the possibility for direct interactions between the magnetic species. This is illustrated in Figure \ref{fig:structure} (b) for the case when the latter occupies the B$^\prime$ sublattice. The dominant exchange path is B$^\prime$ - X - B - X - B$^\prime$, which has previously been described as a super-superexchange type of magnetic interaction \cite{Koo2005, Zhang2015}. Naturally, the choice of the (B, B$^\prime$) pair is paramount to enabling the existence of any significant magnetic interactions in this structure.

\medskip

In this work, we search for the possibility of obtaining magnetic LFHDPs with reasonably high magnetic transition temperatures. We start by examining magnetic exchange parameters (J$_{i}$) in typical LFHDPs, where the species occupying the B sublattice are known to have +1 oxidation state (Ag, Na) while the other are from the 3d series of elements such as Fe and Ni. Then we look at the compounds where B is a non-magnetic element known to exhibit +2 oxidation state while B$^\prime$ is from the 3d series of elements. Finally, we study candidate structures where both B and B$^\prime$ sublattices are populated by 3d$^{1 - 9}$ elements.

\section{\label{sec:methods}Method}
\subsection{\label{sec:candidates} Candidate compounds}
Since the possible magnetic structures span a large chemical space, we proceed with a high-throughput approach to describe the magnetic properties of Lead-free halide double perovskites.

As the first step, we refer to high-throughput databases \cite{Stefano2012, Jain2013} and previous high-throughput studies\cite{Cai2019, Bartel2020} into the structural and thermodynamic stability of LFHDPs and extract 64 candidate compounds (Table S1 \cite{SM2023}) that were predicted to be thermodynamically stable and have one or both B and B' sites populated with a 3d element. We have considered the modified tolerance factor, $\tau$, introduced by Bartel et. al \cite{Bartel2018} as the marker for structural stability of halide double perovskites. For this study, we consider compounds that lie up to 50 meV/atom above the convex hull to account for errors due to insufficient cutoff energies and poor k-point sampling during the generation of these hulls on material databases.\par
It is prudent to note that during the high-throughput studies, only ferromagnetic structures were considered. As we will demonstrate below, many of the compounds in question depart from their ferromagnetic, cubic perovskite structure and assume a lower symmetry structure in an antiferromagnetic, ferromagnetic, or ferrimagnetic configuration (Table S1 \cite{SM2023}). Naturally, taking these effects into account would affect the stability predictions. 

For each of the candidate compounds, we attempt to find their ground state structure within collinear configurations and calculate relevant magnetic exchange interactions. This is followed up by calculating the magnetic transition temperatures via classical Monte Carlo simulations.

\subsection{\label{sec:mag_conf} Magnetic configurations}

\begin{figure*}[!htb]
	\includegraphics[scale = 0.33]{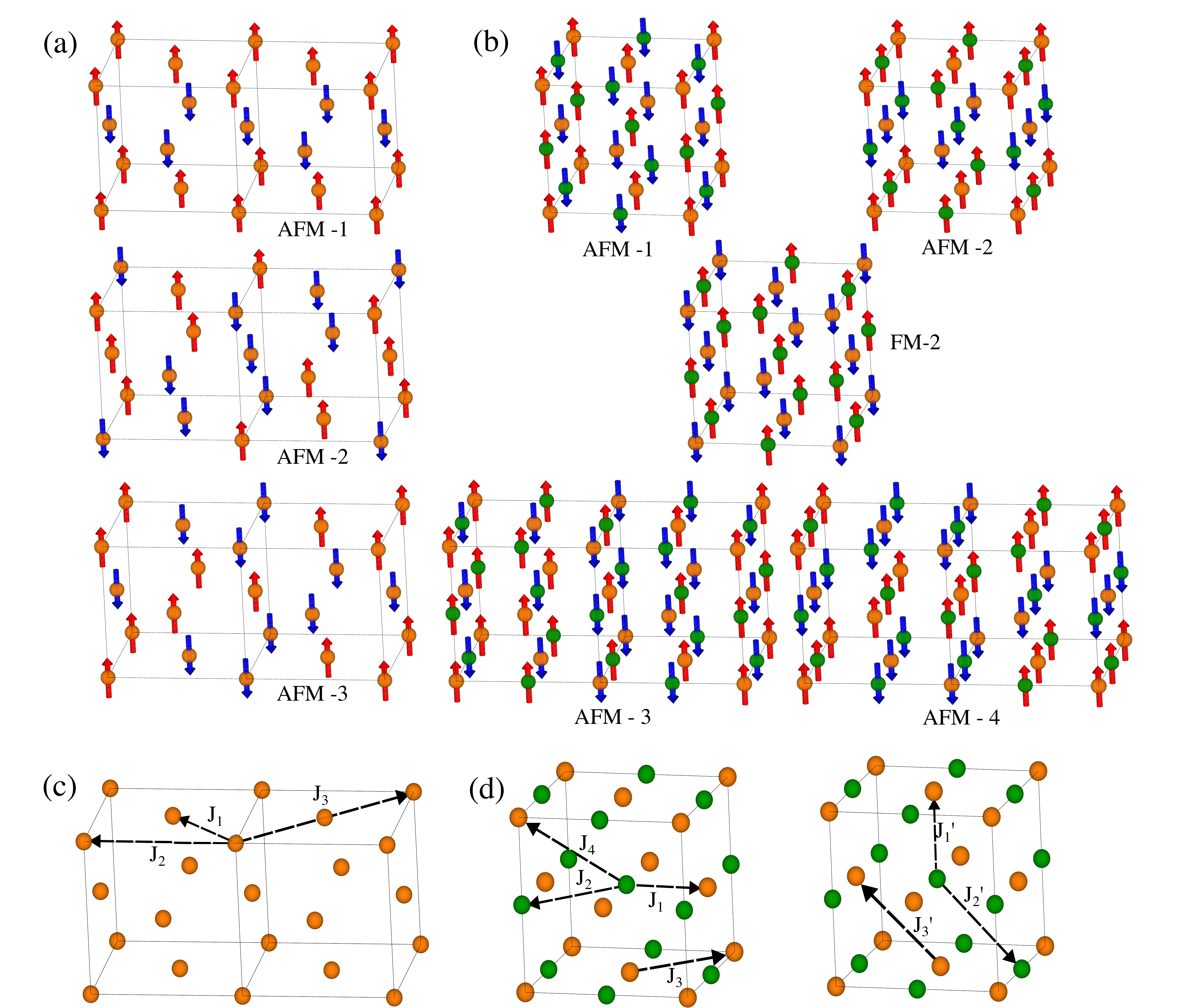}
	\caption{Magnetic configurations for LFHDPs  with (a) B = Non-magnetic ion B$^\prime$ = Magnetic ion (b) B, B$^\prime$ = Magnetic ion. (c) and (d) illustrate the relevant exchange interactions taking tetragonal distortions into account for candidates illustrated in (a) and (b) respectively.}
	\label{fig:magstr}
\end{figure*}

For compounds with a single magnetic element occupying either B or B$^\prime$ site, we consider the typical collinear magnetic configurations allowed in the fcc lattice illustrated in Figure \ref{fig:magstr}(a). For the cases where both B and B$^\prime$ sublattices are occupied by a 3d$^{1}$ - 3d$^{9}$ element, we consider the magnetic configurations illustrated in Figure \ref{fig:magstr}(b).\par

Certainly, it is possible that some of these structures may assume more complex collinear or non-collinear magnetic configurations. However, what we wish to illustrate is the difference in the nature of magnetic interactions with respect to the choice of B and B$^\prime$ ions within this large chemical space. This difference, as we see later, is highlighted quite well using the chosen configurations.

\subsection{\label{sec:exch_int} Magnetic exchange interactions}

The classical Heisenberg Hamiltonian is defined as

\begin{equation}
  H = -\sum_{\langle i j \rangle} J_{ij} \bm{S_i} \cdot \bm{S_j}
  \label{eq:heisenberg}
\end{equation}

Here, $J_{ij}$ represents the exchange interaction between spins at sites $i$ and $j$, and $\bm{S_i}$ and $\bm{S_j}$ are the spin vectors at these sites. The sum is over all pairs of spins, $i \neq j$, in the system.

We calculate the relevant magnetic exchange interactions following the methodology suggested by Fedorova et al. \cite{Fedorova2018}, which is based on the calculation of the total energy of the system with collinear spin alignment when spin states of any two chosen sites, say 1 and 2 within a given unit cell, are modified. Within this method, assuming that the magnetism is fully described by the classical Heisenberg Hamiltonian, Eq. (\ref{eq:heisenberg}), the total energy is:

\begin{equation}
	E = - n \, J_{12} \, \bm{S_1} \cdot \bm{S_2} - \bm{S_1} \cdot \bm{h_1} - \bm{S_2} \cdot \bm{h_2} +  E_{all} + E_0 .
  \label{eq:fedorova}
\end{equation} 
In the first term, n is the number of equivalent bonds with exchange coupling $J_{12}$, $\bm{h_1} =  \sum_{i \neq 1,2} J_{1i} \, \bm {S_i}$, $\bm{h_2} = \sum_{i \neq 1,2} J_{2i} \, \bm{S_i}$. The second term $E_{all} = \sum_{i,j \neq 1,2} J_{ij} \, \bm{S_i} \cdot \bm{S_j}$, and the last term $E_0$ consists of all non-magnetic energy contributions.

In order for the values to be comparable to other methods, we set $|\bm{S_1}| = |\bm{S_2}| = 1$. Then, the relevant exchange interactions between any sites 1 and 2 in the cell can be expressed as

\begin{equation}
	J_{12} = \frac{E_{\uparrow\downarrow} + E_{\downarrow\uparrow} - E_{\uparrow\uparrow} - E_{\downarrow\downarrow}}{4n},
\end{equation}

where $E_{\uparrow\downarrow}$ corresponds to spin-up on site 1 and spin-down on site 2 and so on.

For this study, we calculate the relevant J$_{ij}$s with respect to the calculated ground state magnetic configuration in a 160-atom supercell, which provides good convergence of the results.

In the text following this section, we abbreviate the exchange interaction for the nearest-neighbour species 1 and 2 as J$_1$, the second nearest-neighbour species 1 and 2 as J$_2$ and so on. Similarly, J$_i ^\prime$ refers to the out-of-plane exchange interaction as illustrated in Figure \ref{fig:magstr} (d).

\needspace{0cm}

\section{\label{sec:comp_details}Computational details}

The density functional theory (DFT) computations were performed with the Vienna ab initio simulation package (VASP) \cite{Kresse1996, Kresse1996_2, Kresse1999} using the projector augmented - wave (PAW) method \cite{Blochl1994}  with a plane wave cutoff ENCUT set to 1.5 times the maximum ENMAX in the employed PAW potentials. The recommended VASP PAW potentials for each element \cite{vasp_paw} were used for this study. We performed full structural relaxation with a convergence criterion of 10$^{-8}$ eV for total energy in self-consistent field cycles and $<$ 1 meV/\AA{} for atomic forces.  A k-point density of 0.15 \AA{}$^{-1}$ for structural relaxations and 0.10 \AA{}$^{-1}$ for accurate total energy calculations was used throughout via the KSPACING tag in VASP.\par

The PBEsol \cite{Perdew2008} exchange-correlation functional with Hubbard U correction was used for geometry optimization as well as the calculation of the density of states. We choose the value of the effective on-site correction, U$_{eff}$ \cite{Dudarev1998}, to be a uniform value of 3 eV in this study, which has recently been shown to produce results comparable to hybrid functionals when reproducing the electronic and magnetic properties of this class of materials \cite{Klarbring2023}.

\medskip

The Monte Carlo (MC) simulations were performed with the UppASD \cite{Skubic2008} code using the exchange interactions obtained via DFT calculations. Starting from a temperature of 1000 K and moments aligned in a ferromagnetic configuration, a temperature sweep was performed down to 1~K in order to observe the magnetic phase transition by identifying the temperature at which the curves for fourth-order reduced Binder's cumulant $U^*$ \cite{Binder1981, Binder1981_2, Binder2009} calculated at different system sizes intersect. Due to low ordering temperatures in many cases, the Heat Bath Monte Carlo algorithm, as implemented in the UppASD code, was used to effectively sample the configuration space.

\medskip

Details on chemical bonding were obtained from a Crystal Orbital Hamiltonian Population (COHP) \cite{Dronskowski1993} analysis as implemented in the LOBSTER package \cite{LOBSTER1, LOBSTER2, LOBSTER3}.

\section{\label{sec:results}Results}

\begin{figure*}[!htb]
	\includegraphics[scale = 0.968]{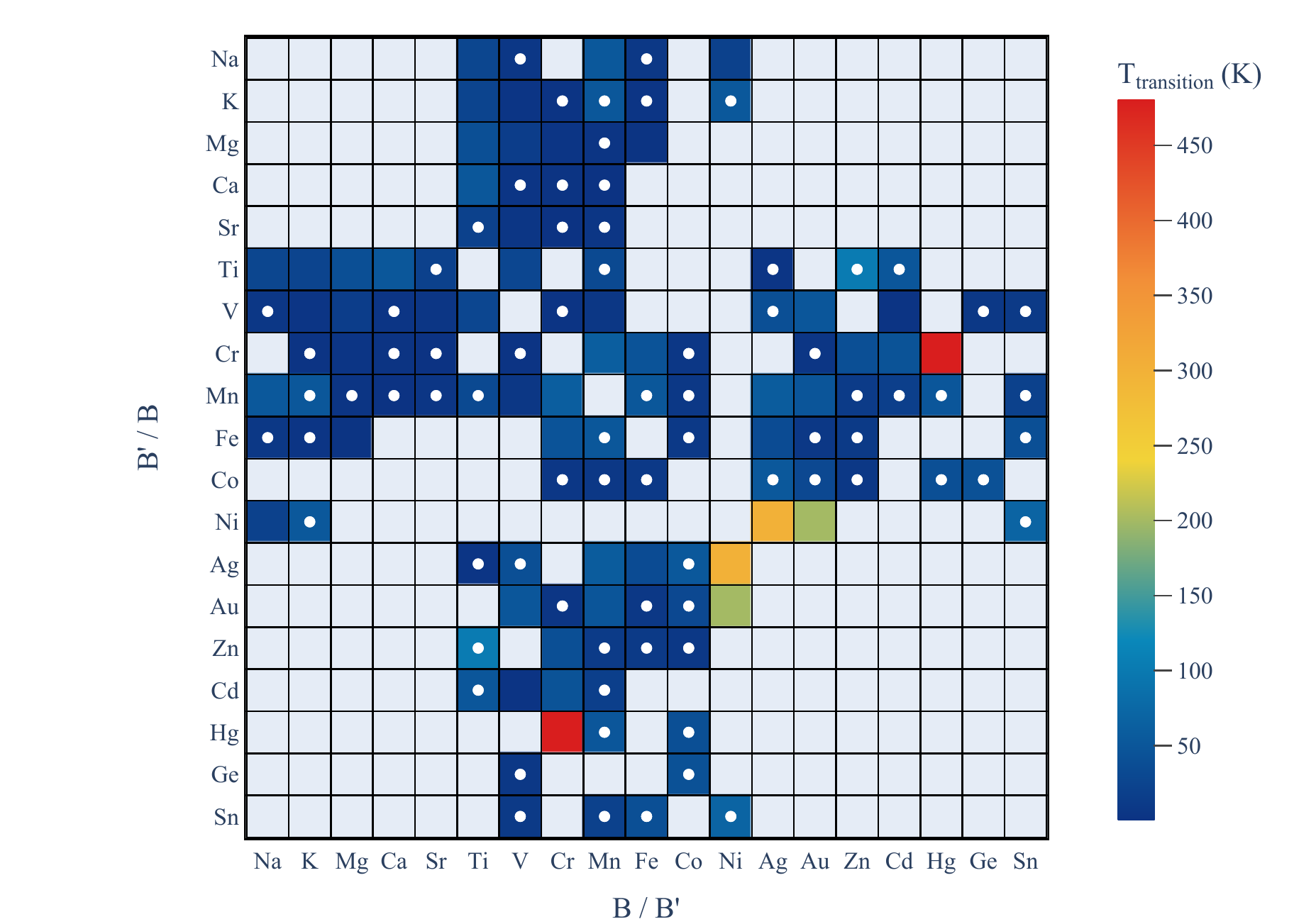}
  \caption {Magnetic transition temperatures for thermodynamically stable LFHDPs, $Cs_2 B B^\prime Cl_6$.
  The grid squares with white dots represent Néel temperatures while those without represent Curie temperatures. Empty grid squares indicate structurally or thermodynamically unstable compositions ($E_{hull} >$  0.050 eV/atom). Elements on both axes are grouped by their occurrence among Alkali metals,  Alkaline Earth metals, Transition Metals, Triels, and Tetrels in the periodic table, respectively, and then by increasing atomic numbers within each group.}
	\label{fig:transition_temp} 
\end{figure*}

\begin{figure}[htbp!]
	\includegraphics[scale = 0.5]{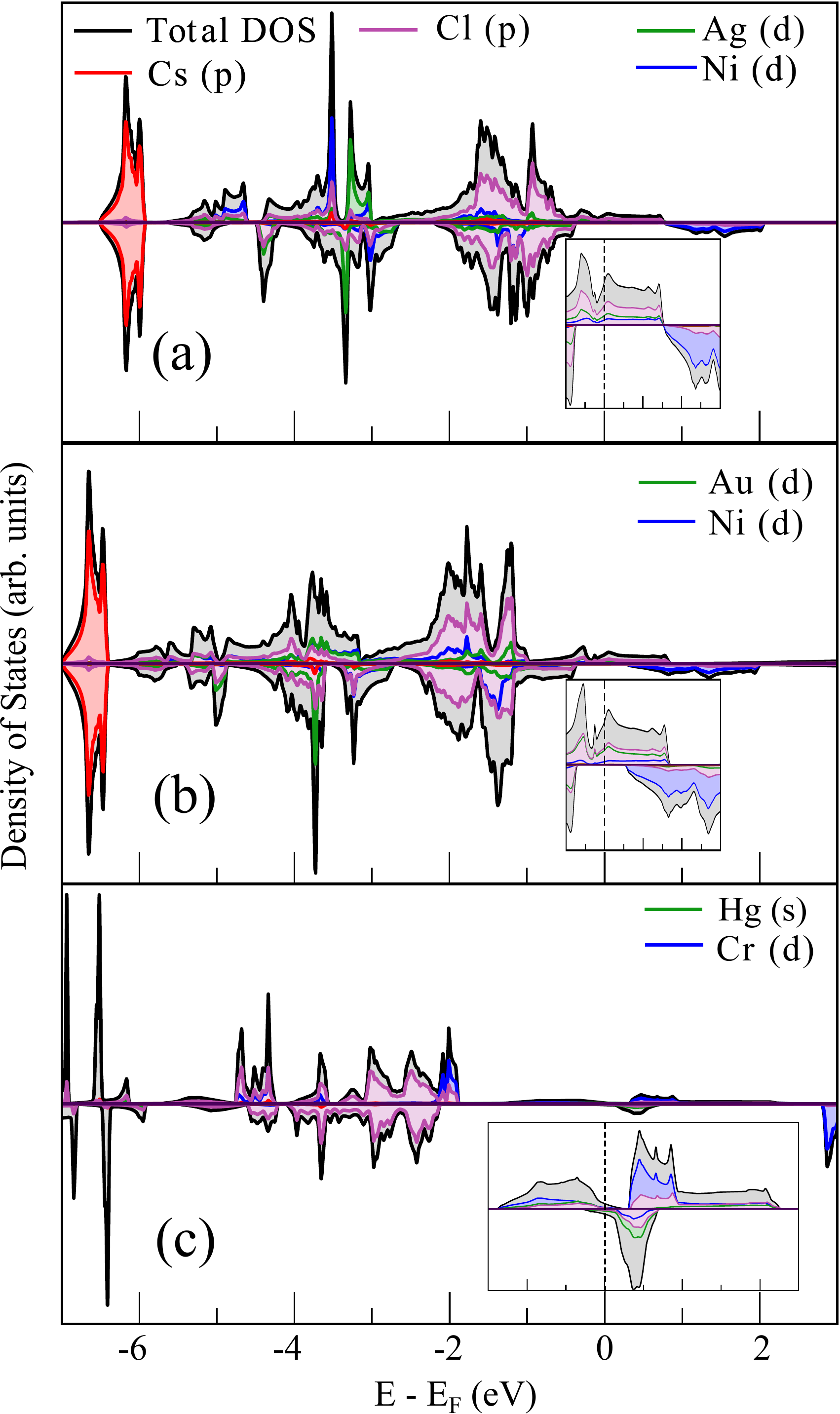}
	\caption {Density of States for $Cs_2BB^\prime Cl_6$, B $\in$ \{Ag, Au, Hg\}, B$^\prime$ $\in$ \{Ni, Cr\}. The Fermi level is set to zero.}
	\label{fig:DOS_ferro}
\end{figure}

\begin{figure*}[!htb]
	\centering
	\resizebox{0.725\textwidth}{!}{\includegraphics[scale = 1.0]{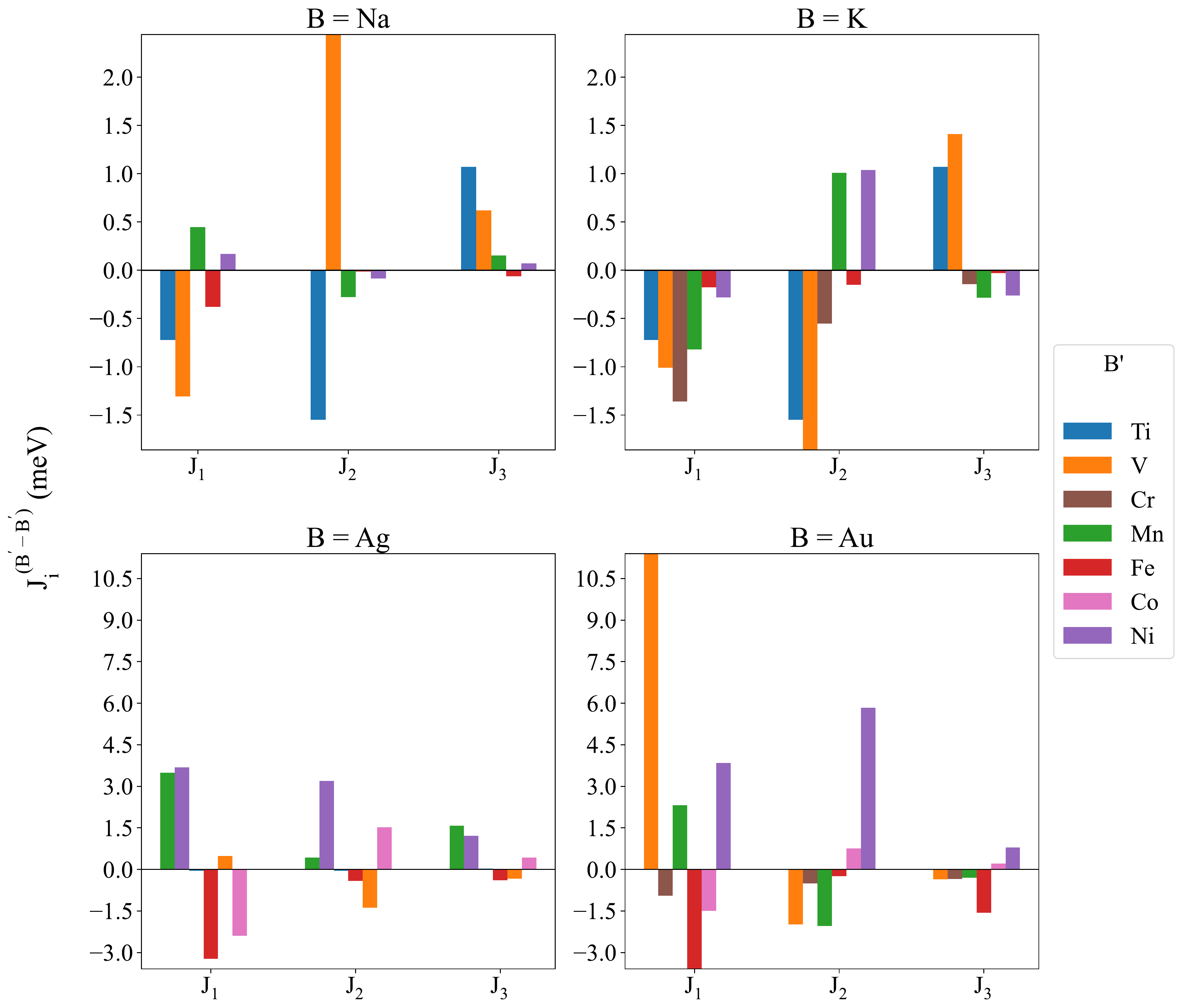}}
	\caption{J$_{i}$ for thermodynamically stable LFHDPs $Cs_2B B' Cl_6$; B $\in$ (Ag, Au, K, Na) and B' $\in$ 3d$^{1 - 9}$}
	\label{fig:J_plus1}
\end{figure*}

\begin{figure*}[!htb]
	\centering
	\resizebox{\textwidth}{!}{\includegraphics[scale = 1.0]{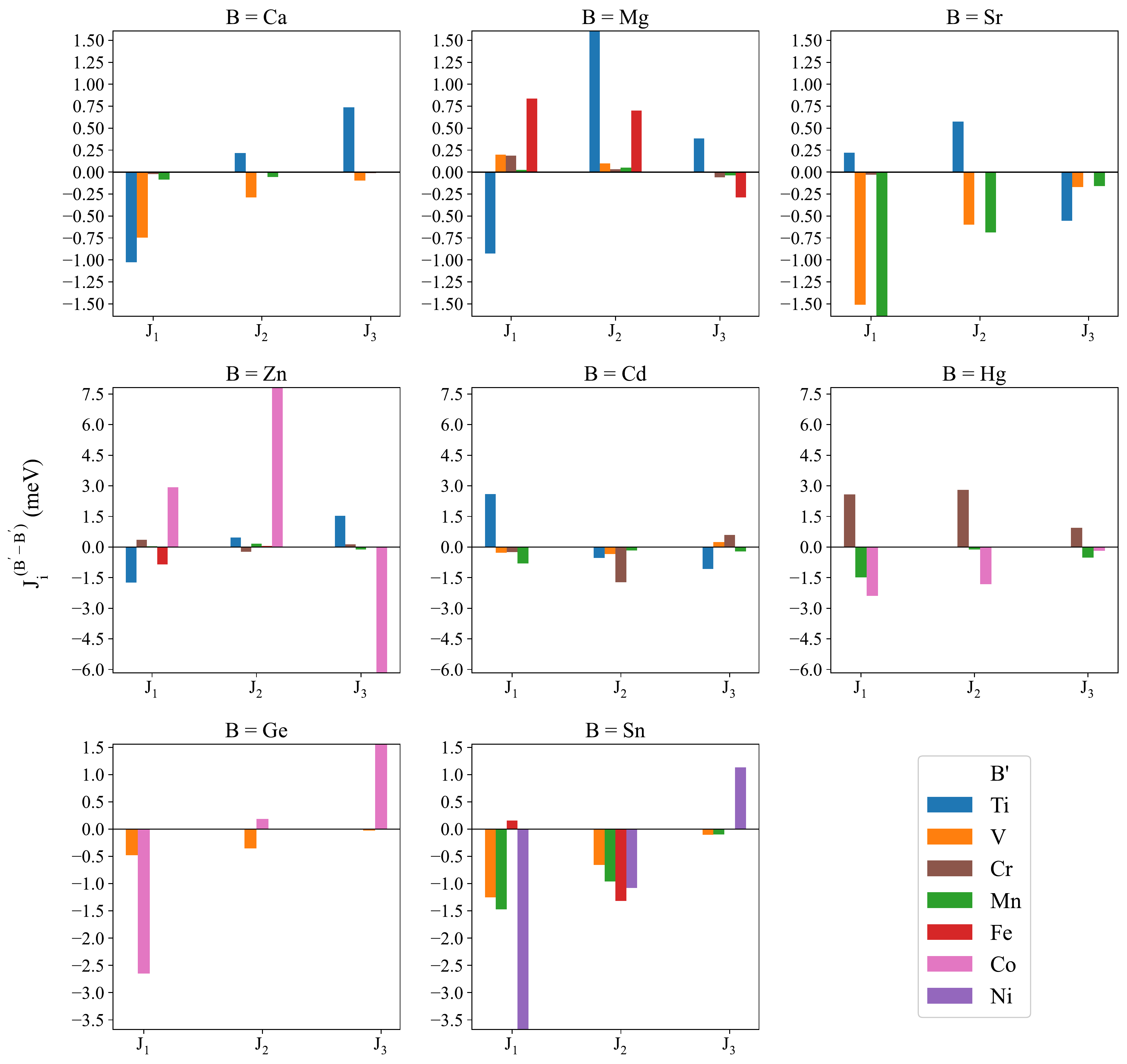}}
	\caption {$J_{i}$ for thermodynamically stable LFHDPs $Cs_2B B' Cl_6$; B $\in$ \{Mg, Ca, Sr, Zn, Cd, Hg, Ge, Sn\} and $B^\prime \in 3d^{1 - 9}$}
	\label{fig:J_plus2}
\end{figure*}

Figure \ref{fig:transition_temp} illustrates the magnetic transition temperatures, as predicted from MC simulations, for the LFHDPs that are the focus of this study. While we predict the majority of the compounds to be magnets with low transition temperatures, this study reveals a few noteworthy exceptions in $Cs_2AgNiCl_6$, $Cs_2AuNiCl_6$, and $Cs_2HgCrCl_6$.\par

These compounds stand out due to much higher magnetic transition temperatures in comparison to other double perovskites with similar chemistry. In the case of $Cs_2AgNiCl_6$, we find T$_{transition}$ to be $\sim$ 190 K, while for $Cs_2AuNiCl_6$ it is predicted to be slightly higher at $\sim$ 240 K. Both of these LFHPs undergo tetragonal distortion. For $Cs_2HgCrCl_6$, which retains its cubic phase, T$_{transition}$ is predicted to be much higher at $\sim$ 430 Kelvin. We predict all three of these compounds to be ferromagnets. 

\medskip

Another interesting feature that is observed in the electronic structure of $Cs_2BNiCl_6$ (B = Ag, Au) is half-metallicity \cite{DeGroot1983, Katsnelson2008}, as illustrated in the insets of Figures \ref{fig:DOS_ferro}(a) and (b) respectively. Here one of the spin channels stays metallic , with the other exhibiting insulating behaviour. This leads to 100\% spin polarization at the Fermi level (E$_{F}$). As half-metals naturally produce highly spin-polarized currents, they are prime candidates for spintronic devices \cite{Igor2004}. 

\medskip

We note that for the case when B, B$^ \prime$ = \{Ag, Ni\}, where ordered magnetism survives up to T$_c$ $\sim$ 190 K, the calculated gap in the insulating channel is 1.4 eV which is larger than some known half-metallic ternary \cite{Jourdan2014} and quaternary \cite{Kundu2017} Heusler alloys that have been proposed for use in spintronics. For B, B$^\prime$ = \{Au, Ni\}, where the ferromagnetic to paramagnetic transition is expected to be at T$_c$ $\sim$ 240K, this gap is nearly half of that, at 0.8 eV. A large gap, and more importantly the unavailability of electronic states near the Fermi level in the insulating channel contributes to the robustness of the half-metallic nature \cite{Katsnelson2008} towards the influence of defects, impurities and temperature effects in these compounds.

\medskip

Having delineated the intriguing feature of half-metallicity in these systems and its implications for spintronics, we now turn to the aspect that provides fundamental understanding of their magnetic behaviour. A multitude of magnetic properties - including the magnetic transition temperatures, susceptibility, coercivity, remanence, and the critical field for magnetic phase transitions - are primarily determined by the strength and nature of exchange interactions within the material.

\medskip

In particular, the main focus of this work, the magnetic transition temperatures, are intimately tied to the robustness of exchange interactions, resisting thermal perturbations at higher temperatures and subsequently enhancing the transition temperature. Further, properties like magnetic susceptibility and coercivity are substantially modulated by these interactions.\par

Therefore, unravelling the intricacies within the mechanism of exchange interactions is paramount to a comprehensive understanding and potential manipulation of the magnetic response in these double perovskite systems. With the following analysis, we aim to shed light on these facets.

\medskip

Discussing experimentally known alloys $Cs_2NaFeCl_6$ \cite{Zhang2023, Armer2023} and $Cs_2AgFeCl_6$ \cite{Yin2020} first, which belong to the B, B$^\prime$ $\in$ (+1, +3) space, we observe that the exchange interactions when B = Ag are an order of magnitude larger than when B = Na (Fig. \ref{fig:J_plus1}). This is due to the interactions along the path B$^\prime$ - X - B - X - B$^\prime$ being mediated by a p - d hybridized orbital, like Ag(d) - Cl(p) for the former as opposed to when there are no p - d interactions and the bonds between the B$^+$ species and the ligand have mainly ionic character, as in the latter case. In addition, for both types of compounds, the exchange interactions beyond the first shell are quite weak in comparison to J$_1$.\par

Expectedly, the same trend is observed for their analogues, $Cs_2KFeCl_6$ and $Cs_2AuFeCl_6$.

\medskip

For the two compounds with standout T$_{transition}$ when B, B$^\prime$ $\in$ \{+1, +3\}, $Cs_2AgNiCl_6$ and $Cs_2AuNiCl_6$, the ferromagnetic configuration departs from the cubic perovskite symmetry due to tetragonal distortion, and is more stable than the calculated antiferromagnetic configurations which also undergo distortion upon relaxation. We attribute the distortion mainly to Ni$^{+3}$ having a Jahn - Teller (JT) active 3d$^7$ configuration as it occurs regardless of how neighbouring spins are aligned relative to each other.\par

In addition, the exchange interactions do not decay as rapidly with the distance between magnetic ions compared to other cases, which is the likely factor that drives up the magnetic transition temperatures.

\medskip

Apart from these exceptions, we predict all +1, +3 combinations of elements on B, B$^\prime$ to exhibit T$_{transition}$ around or below 50 Kelvin as illustrated in Figure \ref{fig:transition_temp}. It is also worthwhile to note that for some combinations of B, B$^\prime$ elements, while the strength of magnetic interactions does not decay as quickly with distance, ordering temperatures are still low. This is attributed to the alternation of ferromagnetic and antiferromagnetic interactions with increasing distance. While the presence of ferromagnetic interactions promotes long range order, competing antiferromagnetic interactions introduce magnetic frustration in the system, thus lowering ordering temperatures and resulting in weaker magnets.

\medskip

Figure \ref{fig:J_plus2} shows the calculated values of J$_{i}$ for different LFHDPs where B $\in$ \{Mg, Ca, Sr, Zn, Cd, Hg, Ge, Sn\}. Akin to when B, B$^\prime$ = (+1, +3) element, the majority of magnetic exchange interactions, when both B and B$^\prime$ elements are populated by species in +2 oxidation state, are also small and decay rapidly beyond the first coordination shell. This is again, in part, owed to large interatomic distance $\approx$ 7 \AA{} between nearest-neighbour magnetic ions. For the case of the exception observed when B, B$^\prime$ = \{Hg, Cr\}, the exchange interactions across the first three shells are all ferromagnetic and do not decay as quickly with interatomic distance. Interestingly, the ferromagnetic configuration retains the cubic symmetry while antiferromagnetic configurations undergo significant tetragonal distortion, indicating the effect of spin-lattice coupling.

\medskip

As shown in Figure \ref{fig:DOS_ferro} (c) (inset), rather unconventional Cr(t$_{2g}$) - Hg(s) hybridized states are observed around the Fermi level, pointing to a peculiar picture, where, in addition to the typical B - X interactions mediating the exchange, a direct B$^\prime$ - B interaction might also provide a significant contribution. The calculated ferromagnetic configuration reveals a magnetic moment of $\approx$ 3.4 $\mu_{B}$/ atom, indicating that the Cr$^{+2}$ ion is in high spin d$^4$ configuration.

\medskip

It is known that d-block elements in the 3d series can exhibit various oxidation states as a function of their chemical environment, which makes it a possibility for LFHDP systems to accommodate two 3d elements on the (B, B$^\prime$) sites. Coupled with the fact that when both B and B' sites are populated with a 3d$^{1-9}$ element, the nearest neighbour distance between two nearest-neighbour magnetic atoms is reduced, it makes for a captivating study where significant magnetic interactions may be expected.

\medskip

\begin{table}[htbp!]
    \centering
        \begin{tabular}{cccccccccc}
            \textbf{B}  & \textbf{B$^\prime$} & \textbf{J$_1$} & \textbf{J$_1^\prime$} & \textbf{J$_2$} & \textbf{J$_2^\prime$} & \textbf{J$_3$} & \textbf{J$_3^\prime$} & \textbf{c/a} \\ \hline
            Ti & Mn &  -7.61  & --  &  -6.94   & --     &  -8.4     &  --      & 1.00 \\
            & V  &  -5.84     &   --    & 13.90      &  --       &  -6.80     &  --      & 1.00 \\ \hline
            V  & Cr &  -11.17     &  -10.11      & 0.03      &    -0.02      & -0.23       &   -0.12   & \textbf{1.06} \\
            & Mn &  1.32     &  --      & -0.33      &   --     &  0.06     &  --   &    1.00 \\ \hline
            Cr & Co & -19.96    &   -16.99     &   -0.07   &  -0.06    & 0.10      &   0.08     &   \textbf{1.05}\\
            & Fe &   9.72    &  8.16     &   0.49    &  0.36      &  -0.53     &  -0.39      &  \textbf{1.03}\\
            & Mn &  7.97     &  6.96     &  0.66     &  0.71      &  -0.25     & -0.21  &   \textbf{1.04}\\ \hline
            Mn & Co &   -10.61    &  --    &  0.72   &   --     &    -1.71  &   --    &    1.01\\
            & Fe & -9.77    &  --     &   -4.92    &  --      &  12.47     &  --      & 0.99\\ \hline
            Fe & Co &  -7.01    &    -8.12   &   0.11  &  0.12      & -0.17     & -0.24    &   \textbf{0.97}\\ \hline
        \end{tabular}%
    \caption{In plane (J$_1$, J$_2$, J$_3$) and out of plane (J$_1^\prime$, J$_2^\prime$, J$_3^\prime$) magnetic exchange parameters, as illustrated in Figure \ref{fig:magstr} (d). Both B and B$^\prime$ atoms belong to the 3d series of elements}
    \label{tab:table1}
\end{table}

The results of this study are summarized in Table \ref{tab:table1} where we can identify three distinct distributions of exchange interactions. For elemental combinations (V, Cr/Mn), (Cr, Co/Fe/Mn), and (Fe, Co), the nearest-neighbour interactions are an order of magnitude stronger than the second and third nearest-neighbour interactions. For (Ti, Mn/V) and (Mn, Co/Fe), the first, second and third nearest neighbour exchange interactions do not decay as quickly and exhibit relatively larger exchange interactions compared to the compounds discussed previously. Unsurprisingly, many of these alloys do deviate from the ideal cubic perovskite structure and end up in a lower symmetry tetragonal structure which is also reflected in the difference between in- and out-of-plane exchange interactions. We highlight significant (c/a $>$ 1.01 or c/a $<$ 0.99) distortions from the ideal perovskite structure in the magnetic ground state. This happens for the following B, B$^\prime$ pairs: (V, Cr), (Cr, Co), (Cr, Fe), (Cr, Mn), and (Fe, Co). The distortions can again be attributed to symmetry breaking via JT-like effects as it is likely that many of these 3d elements exist in JT active d - shell configurations.

\medskip

Even though these compounds were previously predicted to be thermodynamically stable \cite{Bartel2020}, as summarized in Table S1 \cite{SM2023}, we find that, with the exception of (Co, Mn), there are isolated d-bands above the main valence band which indicate electronic instability (Figure S1 \cite{SM2023}) in the structure originating from the fact that one of the d-block species occupying either B or B$^\prime$ sublattices is not in the desired (+2, +2), or (+1, +3) oxidation states. It is worthwhile to note that many vacancy-ordered halide double perovskites have been reported to exist \cite{Maughan2016, Maughan2018, Maughan2019, Jong2019}, making it a possibility for transition metal based perovskites with incompatible oxidation states to adopt similar, perovskite-derived, structures. In addition, the transition metal based vacancy-ordered double perovskite $Cs_2TiX_6$ (X = Cl, Br) \cite{Kavanagh2022} was also reported recently, indicating that Ti might prefer to be in +4 oxidation state in the HDP structure. We also do not expect the 3d transition metals at either site to have +1 oxidation state as it is not commonly observed.

\medskip

Recently, a few related structures with reduced dimensionality have been proposed to accommodate elements with oxidation states incompatible with the typical 3-D double perovskite structure \cite{Ji2021}. While we do not go on to investigate this structural prototype here, we note that it might be possible to have stronger magnetic interactions in such structures and encourage future work to explore this subject.\par

\medskip

As illustrated in Figure \ref{fig:structure} (b), the dominant path for exchange interaction between the first two nearest neighbours (NN) is B$^\prime$ - X - B - X - B$^\prime$ where X - B and B - X bonds have a 90$^{\circ}$ degree and 180$^{\circ}$ angle between them for the first and second NN exchange paths respectively. Naturally, the extent and type of B$^\prime$ - X and B - X interactions in the double perovskite structure affect the strength of this super-superexchange type magnetic interaction.

\medskip

A comparison between magnetic properties of halide double perovskites (DP) can be made against their analogues in oxide double perovskites ($A_2MM^\prime O_6$) as the exchange paths are presumed to be similar \cite{Fang2001}. Here, A $\in$ +2 ions (Ca, Sr) and M, M$^\prime$ are typically metals with oxidation states summing up to +8. We choose a few model compounds in $Sr_2FeM^\prime O_6$ where, like LFHDPs, the   A site is populated with an s-block element. Substituting M = Mo, Re produces ferromagnets with T$_c$ $\sim$ 450 K \cite{kobayashi1998} and T$_c$ $\sim$ 600 K \cite{Sleight1972} respectively, while M = W shows a low N\'eel temperature of $\sim$ 37 K \cite{Kawanka1999}. As far as magnetism is concerned, we highlight some key differences between the halide double perovskites considered in this study and the above-mentioned oxide double perovskites.

\begin{table}[ht!]
    \centering
    \begin{tabular}{cccc|ccccc}
    \rule{0pt}{3ex}\textbf{A} & \textbf{B} & \textbf{B$^\prime$} & \textbf{X} & \textbf{B - B$^\prime$} &
\textbf{B - X} & \textbf{B$^\prime$ - X} & \textbf{B - B} & \textbf{B$^\prime$ -  B$^\prime$} \\ \hline
    \rule{0pt}{3ex}Sr & Mo & Fe & O & 0.09899 & 1.68098 & 0.71787 & 0.01217 & 0.00440\\
    Sr & Re & Fe & O & 0.01632 & 2.33258 & 0.90432 & 0.00079 & 0.00022\\
    Cs & Ag & Fe & Cl & 0.00962 & 0.34040 & 0.75963 & 0.00071 & 0.00023\\
    Cs & Au & Fe & Cl & 0.01392 & 0.19424 & 0.82592 & 0.00097 & 0.00106\\ \hline
    \end{tabular}
    \caption{Comparison of partial (-)ICOHP for B - B$^\prime$, B - X, B$^\prime$ - X, B - B and B$^\prime$ -
B$^\prime$ interactions in representative oxide and halide double perovskite structures. Values are expressed per
pairwise interaction.}
    \label{tab:table2}
\end{table}

\medskip

First, the nearest-neighbour distance between magnetic atoms is shorter in oxide double perovskites, which may enable easier mediation of exchange interactions. This should invariably affect the strength of covalent interactions, key to the mediation of exchange interactions. In addition to this, tilts of MO$_6$ octahedra in oxide DPs are smaller in comparison to the octahedral tilting of B$^\prime$O$_6$ octahedra in halide DPs. Since octahedral tilting reduces the overlap between the orbitals lying in the exchange paths, the weakening of exchange interactions is expected. In order to quantify the extent of covalent bonding between each pair of species, we perform a COHP analysis\cite{Klarbring2023} and calculate the values of -ICOHP, which is the partial -COHP values integrated up to the Fermi level, presented in Table \ref{tab:table2}. While the strength of other interactions is comparable, the B - X and B'- X are much stronger when X = O compared to when X is a halide. This difference could provide insight into why many oxide perovskites are observed to have high magnetic ordering temperatures, while most halide double perovskites do not, as this study reveals.

\medskip

Finally, most halide DPs where the B$^\prime$, B or both sublattices are populated with d-block elements undergo a tetragonal distortion in their predicted, stable ground-state magnetic configuration which is owed to JT-like distortion, spin-lattice coupling or a combination of both, as summarized in Table S1 \cite{SM2023}. Large distortions effectively decrease the number of nearest neighbours for each shell, causing a decrease in significant magnetic interactions as well.

\section{\label{sec:disc}Discussion}

Our findings reveal that HDPs, with either one or both of B, B$^\prime$ sites occupied with a magnetic ion, predominantly form antiferromagnetic configurations with low Néel temperatures. Despite having a similar super-superexchange type mechanism as oxide double perovskites (ODPs), HDPs do not typically exhibit large magnetic ordering temperatures. We attribute this phenomenon to factors including comparatively larger interatomic distances and weaker interactions between the mediating B and X species in HDPs, which leads to less effective exchange coupling and smaller T$_{transition}$ in turn.

\medskip

However, we have identified potential half-metallic LFHDPs in $Cs_2AgNiCl_6$, $Cs_2AuNiCl_6$ and $Cs_2HgCrCl_6$ that have not yet been observed experimentally and may exhibit high Curie temperatures. 100\% spin polarization at the Fermi level combined with relatively high Curie temperatures might make them attractive candidates towards applications in Spintronics \cite{Igor2004}. Aside from these select candidates, ferromagnetic configurations are generally not preferred.\par

\medskip

Alloying at either one or both B and B$^\prime$ sites has emerged as a popular strategy for modulating optical properties in HDPs \cite{Karmakar2018, Wang2022, Ning2020} and a similar stratgy might help drive the nearest-neighbour interactions towards ferromagnetic coupling in order to drive up the magnetic transition temperatures. However, we posit that the effect of sublattice - mixing on the strength of magnetic interactions would likely not be significant enough to promote a substantial change in the transition temperatures. And, even in these cases, the Curie temperatures are expected to remain low due to the inherent limitations posed by the structural factors mentioned above.

\medskip

With the exception of the three candidates mentioned above, in general, this class of compounds may not prove to be promising for magnetic applications beyond 100 Kelvin. Some low-temperature magnetic applications, however, may benefit from the usage of non-toxic and low-cost alternatives to typical magnets, such as halide double perovskites.\par
It has been shown that magnetic materials with low ordering temperatures can exhibit large magnetocaloric effects when alloyed with rare-earth metals\cite{Debye1926, Giauque1927, Giauque1933, Boutahar2017, Wang2013}, which can enhance the efficiency of the cooling cycle in magnetic refrigeration systems. Recently, many rare-earth free candidates for magnetic cooling applications have also been reported \cite{Tan2013, Tegus2002, Krenke2005}. In addition to this, since materials with low magnetic ordering temperatures allow for faster magnetization switching \cite{Katase2016, Wu2022}, the performance of a novel device such as Magnetoresistive Random Access Memory (MRAM) in cryogenic computing could conceivably also benefit from the use of magnetic LFHDPs.

\medskip

It should be noted that while this work has explored the available chemical space of transition metal containing LFHDPs obtained via existing studies into their thermodynamic and structural stability, we have not exhausted the entire chemical space of transition metal containing LFHDPs. It is possible that with a more precise calculation of the convex hull of many other LFHDPs with different species of A, such as Rb$^+$, and X (= Br, I), additional candidates could be identified for further exploration of magnetic properties. This approach would involve considering the different magnetic phases for these candidate LFHDPs, as well as those for competing phases. The underlying mechanisms of magnetic interactions revealed through this study should still hold generally true with a different choice of A and X species in the $A_2B B^\prime X_6$ LFHDP structure. 

\medskip

Finally, although halide double perovskites may not be suitable for high-temperature magnetic applications due to the inherent structural factors that lead to lower transition temperatures, they possess potential for low-temperature applications, offering environmentally friendly and cost-effective alternatives to conventional magnetic materials. Future research could focus on the experimental realization of the proposed ferromagnetic candidates and further exploration of alloying strategies for optimizing magnetic properties in halide double perovskites, while taking into account the underlying structural limitations.

\section{\label{sec:conc}Conclusions}

In this study, we have systematically investigated the magnetic properties of halide double perovskites, focusing on the effect of mediating ions on the strength of exchange interactions and resulting magnetic transition temperatures.
\medskip

The studied LFHDPs show a general propensity towards antiferromagnetic ordering with N\'eel temperatures less than 50 K. However, some ferromagnetic candidates with relatively high Curie temperatures are discovered.
\medskip

While $Cs_2HgCrCl_6$ is predicted to have a Curie temperature of 450K, which is comparable to known oxide double perovskites, $Cs_2AgNiCl_6$ and $Cs_2AuNiCl_6$, with Curie temperatures around 200 K are predicted to be half-metals and proposed for potential use in spintronics. A generally high on-site magnetic moment in the paramagnetic state is also predicted.
\medskip

Via contextualizing the nature of magnetic exchange interactions within the strength and type of bonding interactions, analysed using COHP analysis, we reveal the reasons behind the predicted magnetic properties. 

\section{\label{sec:acknowledge}Acknowledgements}

This work was financially supported by Knut and Alice Wallenberg Foundation (Dnr. KAW 2019.0082). SIS and IAA acknowledge support from the Swedish Government Strategic Research Areas in Materials Science on Functional Materials at Link\"{o}ping University (Faculty Grant SFO-Mat-LiU No. 2009-00971). S.I.S. acknowledges the support from Swedish Research Council (VR) (Project No. 2019-05551) and the ERC (synergy grant FASTCORR project 854843). I.A.A. is a Wallenberg Academy Scholar (Grant No KAW-2018.0194).

The computations were enabled by resources provided by the National Academic Infrastructure for Supercomputing in Sweden (NAISS) and the Swedish National Infrastructure for Computing (SNIC) at the National Supercomputer Center (NSC) and Center for High Performance Computing (PDC), partially funded by the Swedish Research Council through Grant Agreements No. 2022-06725 and No. 2018-05973.


%

\end{document}